%% file: paper.tex
\documentclass[amsmath,amssymb,twocolumn,superscriptaddress]{revtex4-1}

\usepackage{graphicx}
\usepackage{dcolumn}
\usepackage{bm}
\usepackage[detect-all]{siunitx}
\usepackage{hyperref}
\usepackage{xr}
\begin{document}

\title{Bayesian determination of the effect of a deep eutectic solvent on the
structure of lipid monolayers}

\author{A.~R.~McCluskey}
\thanks{A.R.M. and A.S.-F. contributed equally to this work}
\email{a.r.mccluskey@bath.ac.uk/andrew.mccluskey@diamond.ac.uk}
\affiliation{Department of Chemistry, University of Bath, Claverton Down,
Bath, BA2 7AY, UK}
\affiliation{Diamond Light Source, Harwell Campus, Didcot, OX11 0DE, UK}

\author{A.~Sanchez-Fernandez}
\thanks{A.R.M. and A.S.-F. contributed equally to this work}
\altaffiliation[Present address: ]{Department of Food Technology, Lund
University, SE-211 00 Lund, Sweden.}
\affiliation{Department of Chemistry, University of Bath, Claverton Down,
Bath, BA2 7AY, UK}
\affiliation{European Spallation Source, SE-211 00 Lund, Sweden}

\author{K.~J.~Edler}
\affiliation{Department of Chemistry, University of Bath, Claverton Down,
Bath, BA2 7AY, UK}

\author{S.~C.~Parker}
\affiliation{Department of Chemistry, University of Bath, Claverton Down,
Bath, BA2 7AY, UK}

\author{A.~J.~Jackson}
\affiliation{European Spallation Source, SE-211 00 Lund, Sweden}
\affiliation{Department of Physical Chemistry, Lund University, SE-211 00
Lund, Sweden}

\author{R.~A.~Campbell}
\affiliation{Division of Pharmacy and Optometry, University of Manchester,
Manchester, UK}
\affiliation{Institut Laue-Langevin, 71 avenue des Martyrs, 38000, Grenoble,
France}

\author{T.~Arnold}
\email{tom.arnold@esss.se}
\affiliation{Department of Chemistry, University of Bath, Claverton Down,
Bath, BA2 7AY, UK}
\affiliation{Diamond Light Source, Harwell Campus, Didcot, OX11 0DE, UK}
\affiliation{European Spallation Source, SE-211 00 Lund, Sweden}
\affiliation{ISIS Neutron and Muon Source, Science and Technology Facilities
Council, Rutherford Appleton Laboratory, Harwell Oxford, Didcot, OX11 0QX,
UK}

\date{\today}

\begin{abstract}
In this work, we present the first example of the self-assembly of phospholipid monolayers at the interface between air and an ionic solvent.
Deep eutectic solvents are a novel class of environmentally friendly, non-aqueous, room temperature liquids with tunable properties, that have wide-ranging potential applications and are capable of promoting the self-assembly of surfactant molecules.
We use a chemically-consistent Bayesian modelling of X-ray and neutron reflectometry measurements to show that these monolayers broadly behave as they do on water.
This method allows for the monolayer structure to be determined, alongside the molecular volumes of the individual monolayer components, without the need for water-specific constraints to be introduced.
Furthermore, using this method we are able to better understand the correlations present between parameters in the analytical model.
This example of a non-aqueous phospholipid monolayer has important implications for the potential uses of these solvents and for our understanding of how biomolecules behave in the absence of water.
  \begin{description}
  \item[Usage]
  Electronic Supplementary Information (ESI) available: All analysis/plotting
  scripts and figure files, allowing for a fully reproducible, and automated,
  analysis workflow for the work presented is available at
  \url{https://github.com/arm61/lipids_at_airdes} (DOI: 10.5281/zenodo.2577796)
  under a CC BY-SA 4.0 license.
  Reduced experimental datasets are available at
  DOI: 10.15125/BATH-00548, under a CC-BY 4.0 license.
  \end{description}
\end{abstract}
\maketitle

\section{Introduction}
Deep eutectic solvents (DES) are green, sustainable liquids that are obtained through the combination of ionic species with compounds that act as hydrogen bond donors, such as sugars, alcohols, amines, and carboxylic acids \cite{smith_deep_2014,dai_natural_2013}.
The resulting extensive hydrogen bonding network is able to stabilise the ionic species and allows the eutectic mixture to remain liquid at room temperature \cite{hammond_liquid_2016,hammond_resilience_2017,araujo_inelastic_2017}.
Through different combinations of the precursor materials, it is possible to tune the solvent's physicochemical properties, such as polarity \cite{pandey_how_2014}, viscosity and surface tension \cite{smith_deep_2014}, network charge \cite{zahn_charge_2016}, and hydrophobicity \cite{ribeiro_menthol-based_2015,van_osch_hydrophobic_2015}.
Recently DES have also been shown to exhibit a ``solvophobic'' effect through the promotion of surfactant micelle formation \cite{sanchez-fernandez_micellization_2016,arnold_surfactant_2015,hsieh_micelle_2018,banjare_self-assembly_2018}, phospholipid bilayer formation \cite{bryant_effect_2017,bryant_spontaneous_2016,gutierrez_freeze-drying_2009}, and the ability to stabilise non-ionic polymer \cite{sapir_properties_2016} and protein conformations \cite{sanchez-fernandez_protein_2017}.

Phospholipid monolayers at the air/water interface have been widely studied as simplistic models for biological membranes.
As such, they have been used to gain insight into many biological processes that are technologically and medically relevant.
For example, investigations at the air/salt-water interface have identified the importance that interactions between charged phospholipid heads and ions present in solution have on the structure, monomer packing and stability of the monolayer \cite{mohwald_phospholipid_1990,kewalramani_effects_2010}.
However, the native environment for lipids \emph{in-vivo} is far from a simple aqueous solution.
In fact, it has been suggested \cite{dai_natural_2013,hammond_resilience_2017} that DES might form within the crowded cellular environment and could assist in solubilizing biological species in an intermediate environment between that of the hydrophobic phospholipid tails and highly polar water-rich regions, thereby assisting survival under extreme conditions such as freezing temperatures or drought where the water content of cells is restricted.
Despite the broad interest in these systems, the presence of stable phospholipid monolayers at the interface between air and an ionic solvent media has not been previously reported, to the best of the authors' knowledge.

The study of lipid monolayers on a non-aqueous solvent like DES is relevant to some potential new scientific areas or technological applications.
For example, it may be possible to study the interactions of proteins with model membranes in a zero or low water environment, or even at low temperatures where aqueous based model membranes are not practical.
To enable such experiments we must first understand how lipid monolayers are affected by a non-aqueous environment.
Published studies covering such non-aqueous systems are, however, rather scarce mirroring the availability of solvents that show a solvophobic effect comparable to water.
In the absence of water, formamide has been reported to support the formation of lipid monolayers \cite{graner_phospholipidic_1995, weinbach_self-assembled_1993}.
Monolayers have also been observed at the air-mercury interface \cite{magnussen_self-assembly_1996, kraack_structure_2002}, although in this case, the similarity to water is somewhat limited.
We have recently shown that self-assembly of surfactant monolayers is possible for DES with behaviour that is much more comparable to that seen in water \cite{arnold_surfactant_2015, sanchez-fernandez_self-assembly_2018}.

\begin{figure}
\includegraphics[width=0.3\textwidth]{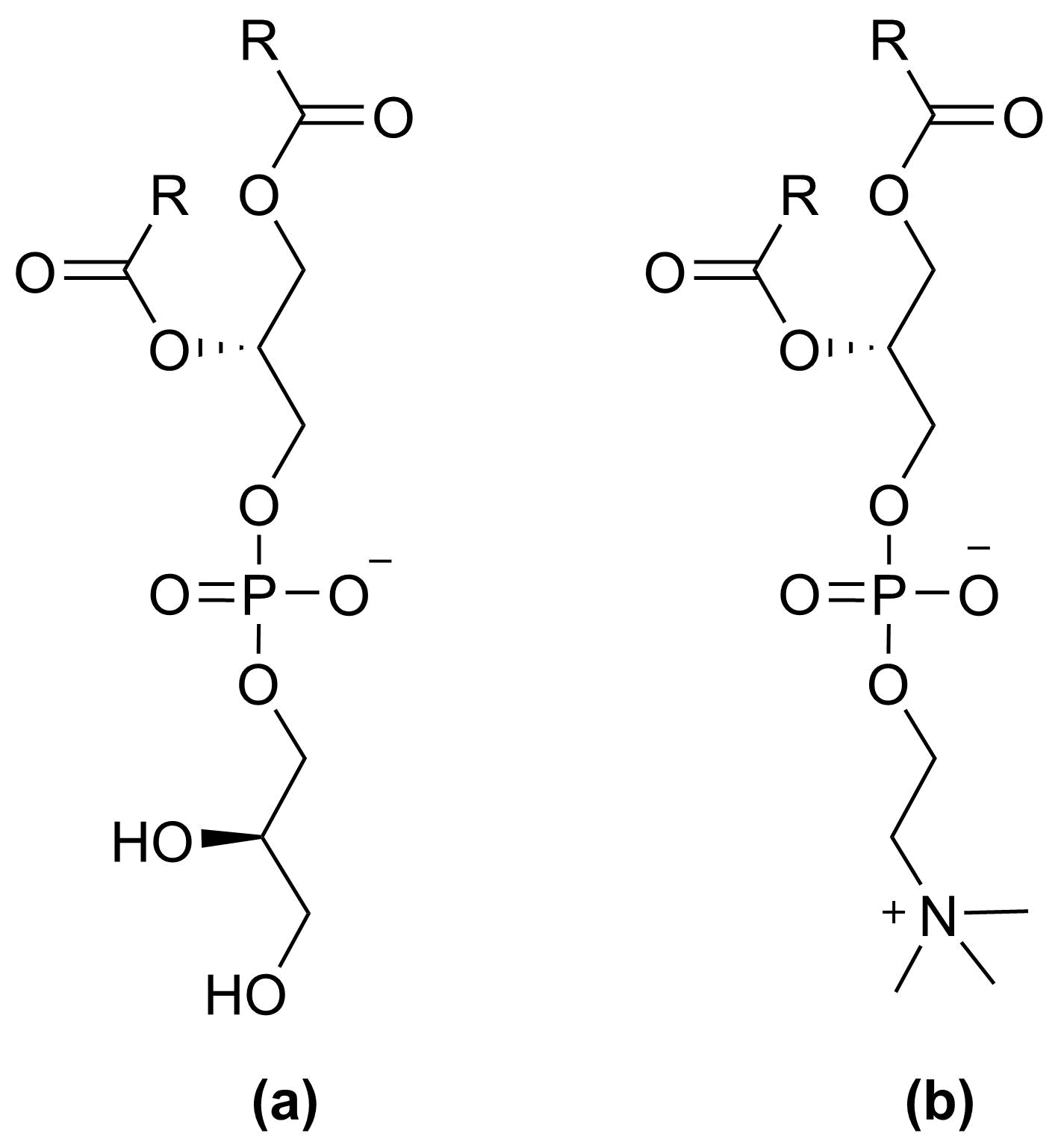}
\caption{\label{fig:heads}\small The two lipid classes with different head
groups compared in this study, where R indicates the hydrocarbon tail;
(a) phosphatidylglycerol (PG), (b) phosphocholine (PC).}
\end{figure}

In this work, we have investigated the structure of phospholipid monolayers at the air-DES interface.
To do this we have used the combination of both X-ray and neutron reflectometry (XRR \& NR), techniques that allow detailed structures to be determined with near-atomic resolution.
In order to interpret our results, we have used a novel analysis method which allows us to sufficiently constrain our model while using Bayesian inference \cite{abeles_sur_1948, parratt_surface_1954} to properly quantify the errors and correlations inherent in such models of reflectometry data.
Specifically, we have constrained the model to ensure that it is chemically self-consistent across multiple measurements at various surface pressures.
Four different phospholipids; 1,2-dipalmitoyl-sn-glycero-3-phosphocholine (DPPC), 1,2-dimyristoyl-sn-glycero-3-phosphocholine (DMPC),  1,2-dilauroyl-sn-glycero-3-phosphocholine (DLPC) and 1,2-dimyristoyl-sn-glycero-3-phospho-(1'-rac-glycerol) (sodium salt) (DMPG), were studied at the interface between a 1:2 mixture of choline chloride:glycerol and air.
This has allowed a comparison of two chemically distinct phospholipid head components, the zwitterionic phosphocholine (PC) and the anionic phosphoglycerol (PG) to be probed in this ionic solvent (Figure \ref{fig:heads}).
In addition, the effect of the tail chain length in the assembly of the Langmuir monolayer was explored by comparing molecules with progressively increasing tail length (12, 14 and 16 carbon atoms in each of the two hydrocarbon chains in each molecule).
Our study is inevitably limited in extent due to the limited availability of X-ray and neutron beamtime together with the prohibitive expense of deuterated chemicals. However, we believe that this work both hints at interesting differences in behaviour for phospholipids monolayers on DES and demonstrates how our approach to fitting can provide a quantitative understanding of this behaviour.

Recent developments in computational resources and software have enabled powerful methodologies and algorithms to be harnessed by those from non-expert backgrounds.
This has benefitted significantly from open-source software projects such as the Python language \cite{van_rossum_python_1995} and the Jupyter notebooks framework \cite{kluyver_jupyter_2016}.
In the area of NR and XRR, the landscape of data-analysis software is diverse, with a range of software packages available from a variety of sources; refnx \cite{nelson_refnx_2019,nelson_refnx_2018-1}, MOTOFIT \cite{nelson_co-refinement_2006}, Rascal \cite{hughes_rascal_2010} Aurore \cite{gerelli_aurore_2016-1}, Refl1D \cite{kienzle_refl1d_2011}, and GenX \cite{bjorck_genx_2007}.

The use of a Python library, such as refnx, enables the implementation of custom models that contain chemically-relevant information as well as the application of probability distribution function (PDF) sampling techniques.
The Python library emcee \cite{foreman-mackey_emcee_2013} allows refnx to access the Goodman \& Weare Affine Invariant Markov chain Monte Carlo (MCMC) Ensemble method \cite{goodman_ensemble_2010}.
This allows the sampling of the high-dimensionality parameter space, relevant in reflectometry analysis, in a Bayesian fashion, where the new samples are generated with consideration of those sampled previously \cite{sivia_data_2006}.
Bayesian inference gives an understanding of the PDF for the fitted parameters and therefore estimations of their inverse uncertainties and inter-parameter correlations.

\section{Experimental Section}
\subsection{Materials}
Choline chloride (\SI{99}{\percent} \%, Sigma-Aldrich) and glycerol (\SI{99}{\percent}, Sigma-Aldrich), and d$_9$-choline chloride (\SI{99}{\percent}, \SI{98}{\percent} D, CK Isotopes) and d$_8$-glycerol (\SI{99}{\percent}, \SI{98}{\percent} D, CK Isotopes) were purchased and used without further purification.
The DES was prepared by mixing the precursors at a 1:2 ratio of choline chloride:glycerol, and heating at \SI{80}{\celsius} until a homogeneous, transparent liquid formed \cite{smith_deep_2014}.
The solvent was equilibrated overnight at \SI{40}{\celsius} and subsequently stored under a dry atmosphere.
Due to the limited availability of the deuterated precursors, a fully protonated subphase (hDES) and a partially deuterated subphase (hdDES) were prepared and used during the neutron reflectometry (NR) experiment.
The partially deuterated subphase was prepared using the following mixtures of precursors: \SI{1}{\mole} of \SI{0.38}{\mole} fraction of h-choline chloride/\SI{0.62}{\mole} fraction of d-choline chloride; and \SI{2}{\mole} of \SI{0.56}{\mole} fraction of h-glycerol/\SI{0.44}{\mole} fraction of d-glycerol.
The deuterated solvent was prepared following the procedure discussed above.

The water content of the DES was determined before and after each experiment by Karl-Fischer titration (Mettler Toledo DL32 Karl-Fischer Coulometer, Aqualine Electrolyte A, Aqualine Catholyte CG A) in order to ensure water presence was kept to a minimum.
Those measurements showed that the water content of the solvent was kept below \SI{0.3}{wt/\percent} during all the experimental procedures presented here, which we assume to be negligible and have to little impact on the characteristics of the DES \cite{hammond_liquid_2016,hammond_resilience_2017}.

DPPC (C$_{16}$ tails, \SI{>99}{\percent}), DMPC (C$_{14}$ tails, \SI{>99}{\percent}), and the sodium salt of DMPG (C$_{14}$ tails, \SI{>99}{\percent}) were supplied by Avanti Polar Lipids and, DLPC (C$_{12}$ tails, \SI{>99}{\percent}) was supplied by Sigma Aldrich and all were used without further purification.
Deuterated versions of DPPC (d$_{62}$-DPPC, \SI{>99}{\percent}, deuterated tails-only) and DMPC (d$_{54}$-DPPC, \SI{>99}{\percent}, deuterated tails-only) were supplied by Avanti Polar Lipids and used without further purification.
These phospholipids were dissolved in chloroform (\SI{0.5}{\milli\gram\per\milli\liter}) at room temperature.
PC indicates the molcule contains a phosphocholine head component, where PG contains a phosphatidylglycerol head component, these are shown in Figure \ref{fig:heads}.

In the XRR experiment, sample preparation was performed in situ using the standard method for the spreading of insoluble monolayers on water: a certain amount of the phospholipid solution was spread onto the liquid surface in order to provide a given surface concentration.
After the evaporation of the chloroform, it is assumed that the resulting system is a solvent subphase with a monolayer of phospholipid at the interface.
Surface concentration was modified by closing and opening the PTFE barriers of a Langmuir trough.
In order to minimise the volumes used in the NR experiment (to keep the cost of deuterated compounds to a manageable level), it was not possible to use a Langmuir trough.
Instead, small Delrin adsorption troughs were used that did not have controllable barriers.
So, although the surface coverage was nominally the same as used in the X-ray studies, the lack of precise control over the surface pressure meant that it was not appropriate to co-refine XRR and NR contrasts together.

\subsection{Methods}
XRR measurements were taken on I07 at Diamond Light Source, at \SI{12.5}{\kilo\electronvolt} photon energy using the double-crystal deflector \cite{arnold_implementation_2012}.
The reflected intensity was measured in a momentum transfer range from \SIrange{0.018}{0.7}{\per\angstrom}.
The data were normalised with respect to the incident beam and the background was measured from off-specular reflection and subsequently subtracted.
Samples were equilibrated for at least one hour and preserved under an helium atmosphere to minimise the adsorption of water by the subphase.
XRR data were collected for each of the lipids, DLPC, DMPC, DPPC and DMPG at four surface pressures (DLPC: \SIlist[list-units = single]{20;25;30;35}{\milli\newton\per\meter}, DMPC: \SIlist[list-units = single]{20;25;30;40}{\milli\newton\per\meter}, DPPC: \SIlist[list-units = single]{15;20;25;30}{\milli\newton\per\meter}, DMPG: \SIlist[list-units = single]{15;20;25;30}{\milli\newton\per\meter}, as measured with an aluminium Wilhelmy plate; measurements were conducted at \SIlist{7;22}{\celsius}.
The aluminium Wilhelmy plate was used over a traditional paper plate due to the low wettability of paper by the DES.

The NR experiments were performed on FIGARO at the Institut Laue-Langevin using the time-of-flight method \cite{campbell_figaro_2011}.
Data at two incident angles of \SI{0.62}{\degree} and \SI{3.8}{\degree} were measured to provide a momentum transfer range from \SIrange{0.005}{0.18}{\per\angstrom}.
Two surface pressures for each system and contrast was measured (DMPC: \SIlist[list-units = single]{20;25}{\milli\newton\per\meter}, DPPC: \SIlist[list-units = single]{15;20}{\milli\newton\per\meter}).
Similar to the X-ray procedure, samples were given enough time to equilibrate (at least two hours), kept under an inert atmosphere, and all measurements were conducted at \SI{22}{\celsius}.

\subsection{Data analysis}
The use of XRR and NR to analyse the structure of phospholipids on the surface of water is well documented \cite{mohwald_phospholipid_1990,kewalramani_effects_2010,bayerl_specular_1990,johnson_structure_1991,clifton_role_2012,helm_phospholipid_1987,daillant_x-ray_1990}.
The models used in the rationalisation of XRR and NR data have varied significantly in numbers of layers present, use of interfacial roughness, and the parameterisation of the physical constraints applied.
Frequently, these physical constraints include the volumes of the phospholipid head and tail components, using values taken from other techniques, such as those shown in Table \ref{tab:water}.
Additionally, a recent evaluation of the applicability of different models for surfactant and phospholipid monolayers from the NR perspective has been published \cite{campbell_structure_2018}, that suggests possible oversights in the modelling of NR data.
\begin{table*}
    \caption{\label{tab:water} Lipid component volumes extracted from different literature sources.
  $V_l$ corresponds to the total lipid volume, MD to molecular dynamics simulation, WAXS to wide-angle X-ray scattering, NB to neutral buoyancy and DVTD to differential vibrating tube densimetry. $^a$
  The values for the head component in Kucerka \emph{et al.} \cite{kucerka_determination_2004}, were taken from Balgav\'{y} \emph{et al.} \cite{balgavy_evaluation_2001}.}
  \begin{ruledtabular}
    \begin{tabular*}{\textwidth}{l|lll|ll|ll|l|l}
    Lipid & DPPC & & & DMPC & & DLPC & & DMPG & POPG \\
    \hline
    Reference & \cite{armen_phospholipid_1998} & \cite{sun_order_1994} &
    \cite{kucerka_determination_2004,balgavy_evaluation_2001}$^a$ & \cite{armen_phospholipid_1998} & \cite{kucerka_determination_2004,balgavy_evaluation_2001}$^a$ & \cite{armen_phospholipid_1998} & \cite{kucerka_determination_2004,balgavy_evaluation_2001}$^a$ & \cite{pan_molecular_2012} &
    \cite{kucerka_scattering_2012} \\
    \hline
    $V_l$/\AA$^3$ & $1287.3\pm25.5$ & $1148\pm2$ & $1264.2\pm32.1$ &
    $1172.5\pm25.1£$ & $1155.4\pm30.0£$ & $1057.7\pm24.7$ & $1046.6\pm28.0$ &
    $1011.4$ & $1203$ \\
    $V_t$/\AA$^3$ & $966.4\pm5.4$ & $829\pm4$ & $924.7\pm17.6$ &
    $851.5\pm5.0£$ & $815.9\pm15.5£$ & $736.8\pm4.6$ & $707.1\pm13.5$ &
    $720.4$ & $914$ \\
    $V_h$/\AA$^3$ & $320.9\pm20.1$ & $319\pm6$ & $339.5\pm14.5$ &
    $320.9\pm20.1$ & $339.5\pm14.5$ & $320.9\pm20.1$ & $339.5\pm14.5$ &
    $291.0$ & $289$ \\
    Method & MD & WAXS & NB & MD & NB & MD & NB & DVTD & MD \\
    T/$^\circ$C & 50 & 24 & 30 & 50 & 30 & 50 & 30 & 20 & 25 \\
    \end{tabular*}
  \end{ruledtabular}
\end{table*}

In Table \ref{tab:water}, there appears to be a general consensus that the component volume for the phosphocholine (PC) head is around \SIrange{320}{360}{\angstrom\cubed}, while the phosphatidylglycerol (PG) head is in the range \SIrange{289}{291}{\angstrom\cubed}.
However, it is not clear that the head component volumes from the literature, that are derived from water-based measurements, will be appropriate for this work, which involves a non-aqueous solvent.
The charged nature of the zwitterionic or anionic lipid heads may have different interactions with the polar, but neutral water as compared to the charged DES components \cite{sanchez-fernandez_self-assembly_2018}, which will affect the phospholipid head component volume.
Therefore, herein we apply a chemically-consistent model that allows for the co-refinement of reflectometry measurements at different surface pressure and makes no assumption of the component volume for the lipid head, $V_h$, or tail, $V_t$.
Instead, these parameters were allowed to vary for each lipid while being constrained to be self-consistent over different surface pressures in the same phase; Liquid-Condensed (LC) for DPPC and Liquid-Expanded (LE) for DMPC, DMPG, and DLPC.
Furthermore, it is known that, on water, increased surface pressure and the associated LE-LC phase transitions lead to a compression of the lipid tail volume \cite{marsh_molecular_2010,small_lateral_1984}, and this compaction has not necessarily been accounted for in the literature \cite{campbell_structure_2018}.
This model avoids this issue by making no assumption about the molecular volumes and only considers surface pressures that we believe to be in the same phase.

Our chemically-consistent model has been implemented in the Python library refnx \cite{nelson_refnx_2019,nelson_refnx_2018-1}.
This software allows for the inclusion of a custom model to be defined, from which parameters feed into the Abel\`{e}s reflectivity model (a model that is widely used to calculate reflectivity \cite{abeles_sur_1948,parratt_surface_1954}).
This custom model, along with a series of Jupyter notebooks showing, in full, the analysis performed, can be found in the ESI and is available under a CC BY-SA 4.0 licence.

This model consists of two layers to define the lipid monolayer; the head layer at the interface with the solvent and the tail layer at the interface with the air.
The head components have a calculated scattering length, $b_h$, (found as a summation of the X-ray or neutron atomic scattering lengths), and a component volume, $V_h$.
These head components make up a layer with a given thickness, $d_h$, and roughness, $\sigma_h$, within which some volume fraction of solvent can intercalate, $\phi_h$.
The tail layer is defined in the same fashion, except that the thickness, $d_t$, is limited such that it may be no greater than the maximum extended length of the lipid tail (the Tanford length, $t_t$ \cite{tanford_hydrophobic_1980}), which is given in Table \ref{tab:invariant}, and that no solvent may intercalate into the layer (e.g. $\phi_t = 0$).
The scattering length density (SLD) of the tail and head layers used in the Abel\`{e}s model can therefore be found as follows,
\begin{equation}
\text{SLD}_i = \frac{b_i}{V_i}(1 - \phi_i) + \text{SLD}_{s}(\phi_i),
\end{equation}
where, $\text{SLD}_{s}$ is the scattering length density of the subphase (DES), and $i$ indicates either the tail or head layer.
To ensure that the number density of head components and pairs of tail components is the same, the following constraint was included in the model \cite{braun_polymers_2017},
\begin{equation}
\label{equ:phih}
\phi_h =  1 - \bigg(\frac{d_tV_h}{V_td_h}\bigg).
\end{equation}
Based on the work of Campbell \emph{et al.} \cite{campbell_structure_2018}, a single value for the interfacial roughness was fitted for all of the interfaces, including the subphase (i.e. $\sigma_h$ = $\sigma_t$ = $\sigma_s$), as there is only a single lipid molecule type in each monolayer.
Therefore, any capillary wave roughness at the air-DES interface is carried equally through the layers.
The interfacial roughness was constrained to be greater than \SI{3.3}{\angstrom} in agreement with previous work \cite{sanchez-fernandez_micellization_2016}.
\begin{table}
    \caption{\label{tab:invariant} The invariant parameters within the chemically-consistent model.
    $^a$Values obtained from the Tanford formula \cite{tanford_hydrophobic_1980}.
  $^b$Values obtained from Sanchez-Fernandez \emph{et al.}\cite{sanchez-fernandez_micellization_2016}.}
  \begin{ruledtabular}
  \begin{tabular*}{0.48\textwidth}{lllll}
        Component & $b_t$/\si{\femto\meter} & $b_h$/\si{\femto\meter} & $t_t$/\si{\angstrom} & $\text{SLD}$/$10^{-6}$\si{\per\angstrom\squared} \\
        \hline
        X-ray & & & & \\
        DLPC & 5073 & 4674 & 15.5$^a$ & -- \\
        DMPC & 5985 & 4674 & 18.0$^a$ & -- \\
        DPPC & 6897 & 4674 & 20.5$^a$ & -- \\
        DMPG & 5985 & 4731 & 18.0$^a$ & --\\
        Air & -- & -- & -- & 0\\
        DES & -- & -- & -- & 10.8$^b$ \\
        \hline
        Neutron & & & & \\
        d$_{54}$-DMPC & 5329.8 & 602.7 & 18.0$^a$ & -- \\
        d$_{62}$-DPPC & 6129.2 & 602.7 & 20.5$^a$ & -- \\
        h-DES & -- & -- & -- & 0.43$^b$  \\
        hd-DES & -- & -- & -- & 3.15$^b$ \\
    \end{tabular*}
  \end{ruledtabular}
\end{table}

In order to justify the use of a single tail volume across many surface pressures, it was necessary to ensure that the lipids remain in the same phase.
On water, this can be demonstrated with a Langmuir isotherm.
However, while we have confidence that the individual surface pressures measured were reliable, we were unable to collect consistent Langmuir isotherm measurements, due to the high viscosity of the DES.
Instead, we have used grazing incidence X-ray diffraction to confirm the
phases of DMPC and DPPC at \SI{30}{\milli\newton\per\meter}.
DPPC was found to be in the LC phase and DMPC in the LE phase at room
temperature for the surface pressures measured (see Section \ref{sec:gixd} in the ESI).
We assume that DMPG and DLPC are also in the LE phase since there is no
reason to believe that the phase behaviour in these systems differs
significantly from DMPC at the same temperature.

In the first of two steps, this custom model was used to co-refine the component volume of the lipid head component, $V_h$, the volume of the tail component, $V_t$, and the head thickness, $d_h$ across XRR measurements at four different surface concentrations.
In keeping with the work of Campbell \emph{et al.} \cite{campbell_structure_2018}, a single value for the head thickness was fitted for each lipid across all surface pressures, as the thickness of the head layer was considered to be dependent on molecular dimensions only, and has been shown to vary little with surface pressure \cite{kewalramani_effects_2010}.
The following parameters were allowed to vary; $d_t$, and $\sigma_{t,h,s}$, independently across the surface pressures, while others, shown in Table \ref{tab:invariant}, were held constant at the values given.
For each co-refinement of four XRR measurements, there were, in total, eleven degrees of freedom in the fitting process.
Throughout all of the analyses, the reflectometry scale factor was allowed to vary freely, while the background was constrained to the intensity of either the largest or second-largest $q$-value.

In the second step, the head and tail component volumes, and head layer thickness determined from XRR were fixed for the refinement of the custom model against the NR measurements.
This approach means that the number of variable parameters to fit the NR data can be reduced to two, namely the thickness of the tail layer, $d_t$, and the interfacial roughness, $\sigma_{t,h,s}$, for the co-refinement of two datasets.
Table \ref{tab:invariant} also gives the details of the scattering lengths and SLDs used as invariant parameters for the NR fitting.

In both cases, the refinement of the custom model to the experimental data involved the transformation of the reflectometry calculated from the model and the data into $Rq^4$ such that the contribution of the Fresnel decay was removed, before using the differential evolution method available to refnx from the scipy library \cite{jones_scipy_2001}, to find the parameters that gave the best fit to the data.
The parameter space was then probed using the MCMC method available through emcee \cite{foreman-mackey_emcee_2013}, which allowed for an estimate of the probability distribution function (PDF) associated with each parameter.
In the MCMC sampling, 200 walkers were used over 1000 iterations, following equilibration of 200 iterations.
The use of MCMC sampling allowed for a Bayesian inference of the PDF for each of the variables and their respective interactions and the Shapiro test to be used to assess if each PDF was normally distributed.
Parameters that were shown to be normally distributed are given with symmetric confidence intervals, while those that failed the Shapiro test are given with asymmetric confidence intervals (\SI{95}{\percent} confidence intervals in both cases).
However, it is important to note that these are not \emph{true} confidence intervals, and account only for the uncertainty present in the data, i.e. they do not account of systematic uncertainty in the measurement that is underrepresented, or unrepresented, in the experimental dataset.

\begin{figure}
    \centering
    \includegraphics[width=0.45\textwidth]{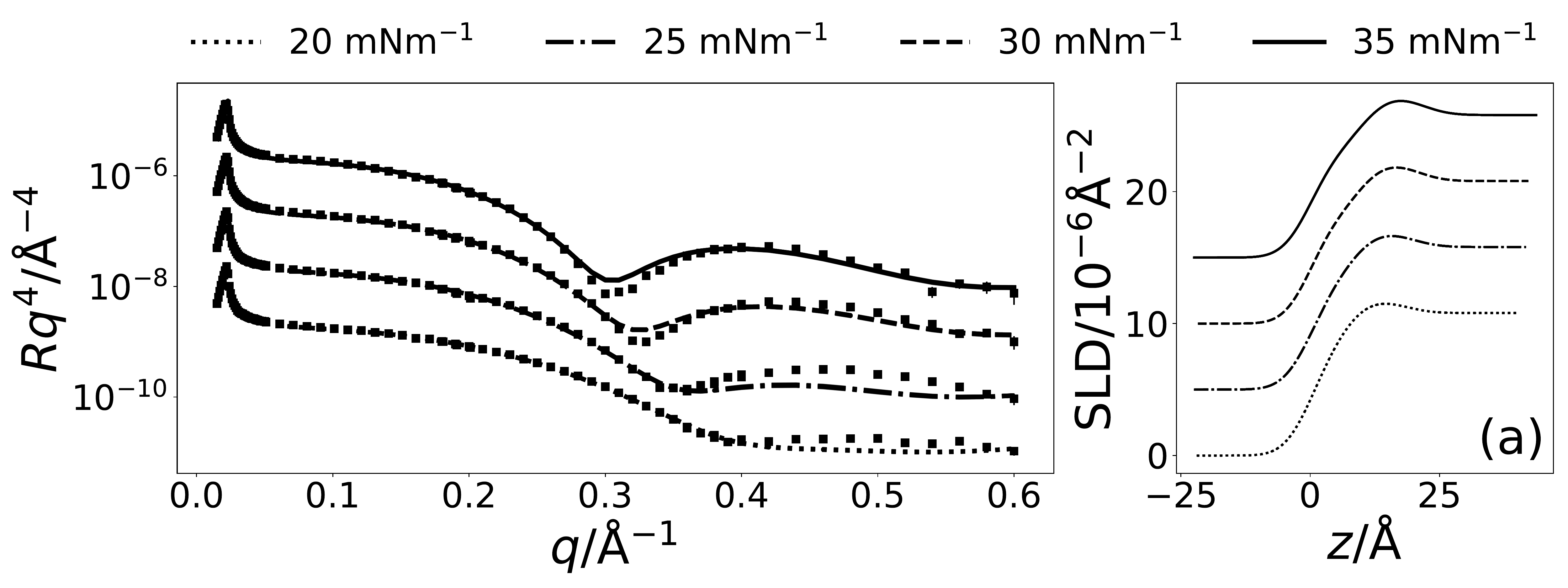}
    \includegraphics[width=0.45\textwidth]{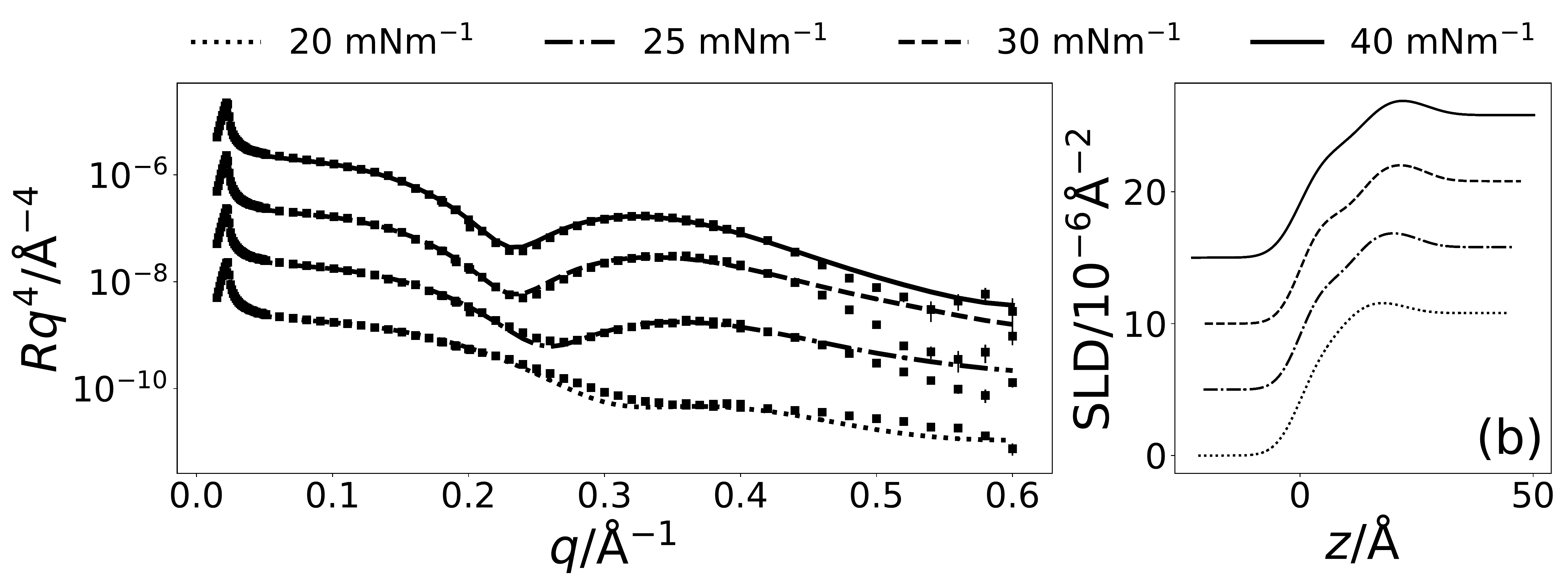}
    \includegraphics[width=0.45\textwidth]{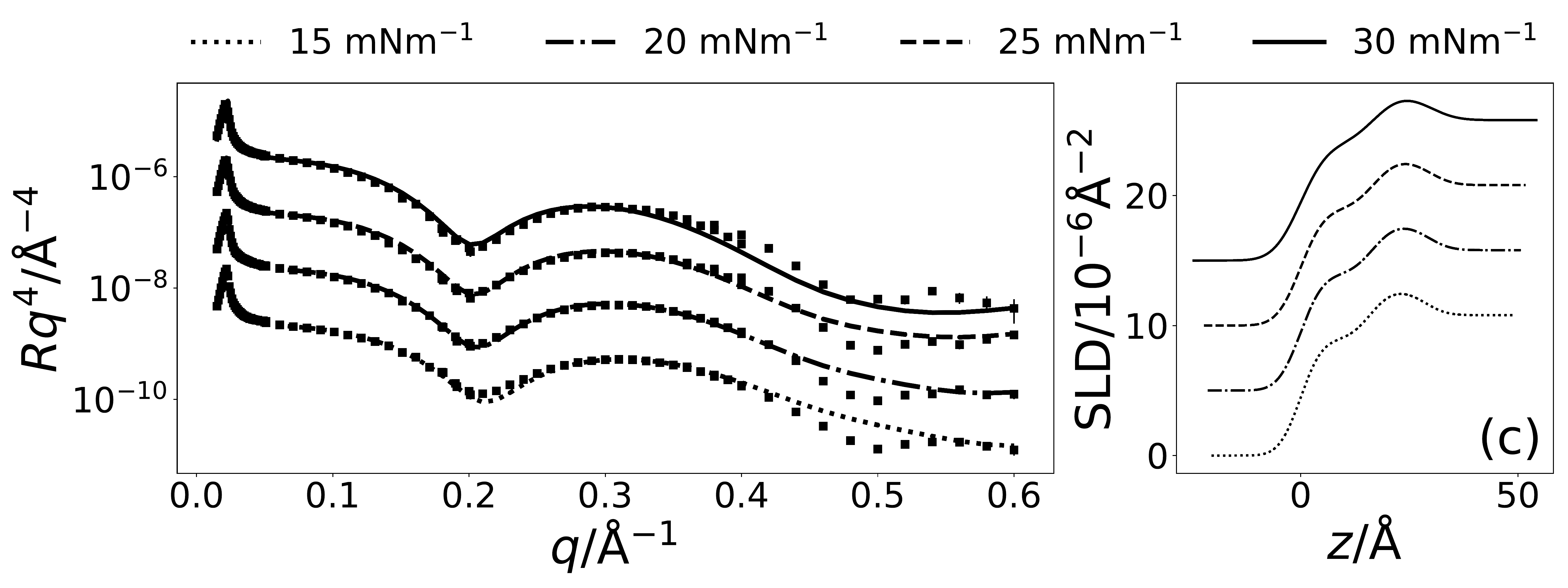}
    \includegraphics[width=0.45\textwidth]{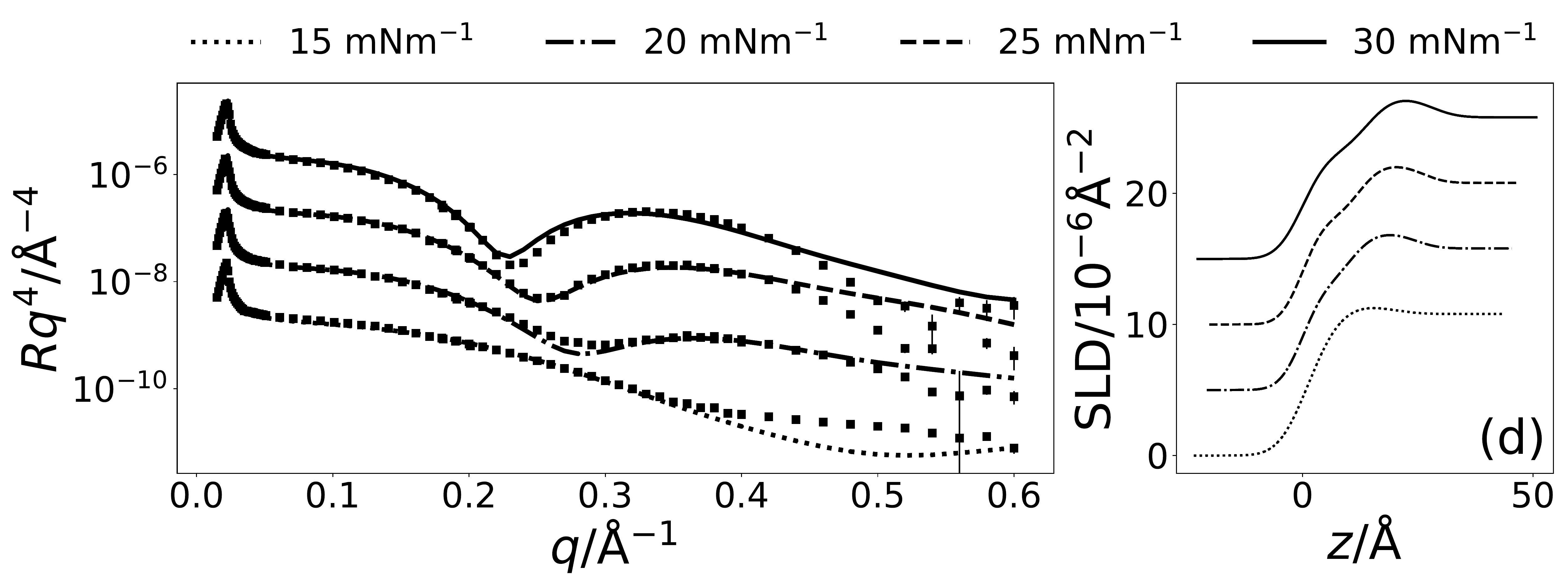}
    \caption{\small The XRR profiles (left) and SLD profiles (right) for each
  of the four lipids; (a) DLPC, (b) DMPC, (c) DPPC, and (d) DMPG, at the four
  measured surface pressures; see legend above each plot. The different
  surface pressure XRR profiles have been offset in the $y$-axis by an order
  of magnitude and SLD profiles offset in the $y$-axis by
  \SI{5e-6}{\per\square\angstrom}, for clarity.}
    \label{fig:lipids}
\end{figure}
\section{Results \& Discussions}
The chemically-consistent model was co-refined across the four surface pressure XRR measurements for each lipid.
Although lipid monolayers on water often reach values above \SI{50}{\milli\newton\per\meter}, signifying a decrease in surface tension to ca. \SI{20}{\milli\newton\per\meter} upon addition of lipid, it was not possible to reach those values for the monolayers on DES \cite{mohwald_phospholipid_1990}.
The surface tension of pure choline chloride:glycerol has been previously reported to be \SI{63.5\pm0.5}{\milli\newton\per\meter} \cite{sanchez-fernandez_micellization_2016}.
This means that a surface pressure of ca. \SI{40}{\milli\newton\per\meter} iin choline chloride:glycerol will also result in a reduction to ca. \SI{20}{\milli\newton\per\meter} in the surface tension.
Interestingly, the absolute values of surface tension in both water and choline chloride:glycerol after addition of lipid are similar.
Therefore, surface pressures of up to \SI{40}{\milli\newton\per\meter} were measured for the monolayers on DES, when possible.
The resulting XRR profiles and associated SLD profiles are shown in Figure \ref{fig:lipids}.
Table \ref{tab:liptab} gives details of all varied parameters for each lipid at \SI{30}{\milli\newton\per\meter}, as well as the details of $\phi_h$ which was determined from Eqn. \ref{equ:phih} (the same details for the other surface pressures are available in Section \ref{sec:xrrpara} of the ESI).
\begin{table*}
    \caption{\label{tab:liptab} The best-fit values, and associated \SI{95}{\percent} confidence intervals for the varying parameters in the XRR models, at the \SI{30}{\milli\newton\per\meter}. The values for $\phi_h$ was obtained from the appropriate use of Eqn. \ref{equ:phih}.}
  \begin{ruledtabular}
    \begin{tabular*}{\textwidth}{ccccc}
    Lipid & DLPC & DMPC & DPPC & DMPG \\
    \hline
    $\sigma$/\si{\angstrom} & \input{dlpc_rough30.txt} & \input{dmpc_rough30.txt} & \input{dppc_rough30.txt} & \input{dmpg_rough30.txt} \\
    $d_t$/\si{\angstrom} & \input{dlpc_tail30.txt} & \input{dmpc_tail30.txt} & \input{dppc_tail30.txt} & \input{dmpg_tail30.txt} \\
    \hline
    $V_t$/\si{\cubic\angstrom} & \input{dlpc_vt.txt} & \input{dmpc_vt.txt} & \input{dppc_vt.txt} & \input{dmpg_vt.txt} \\
    $V_h$/\si{\cubic\angstrom} & \input{dlpc_vh.txt} & \input{dmpc_vh.txt} & \input{dppc_vh.txt} & \input{dmpg_vh.txt} \\
    $d_h$/\si{\angstrom} & \input{dlpc_head.txt} & \input{dmpc_head.txt} & \input{dppc_head.txt} & \input{dmpg_head.txt} \\
    \hline
    $\phi_h$/$\times10^{-2}$ & \input{dlpc_solh30.txt} & \input{dmpc_solh30.txt} & \input{dppc_solh30.txt} & \input{dmpg_solh30.txt} \\
    \end{tabular*}
  \end{ruledtabular}
\end{table*}

\begin{figure}
    \centering
  \includegraphics[width=0.45\textwidth]{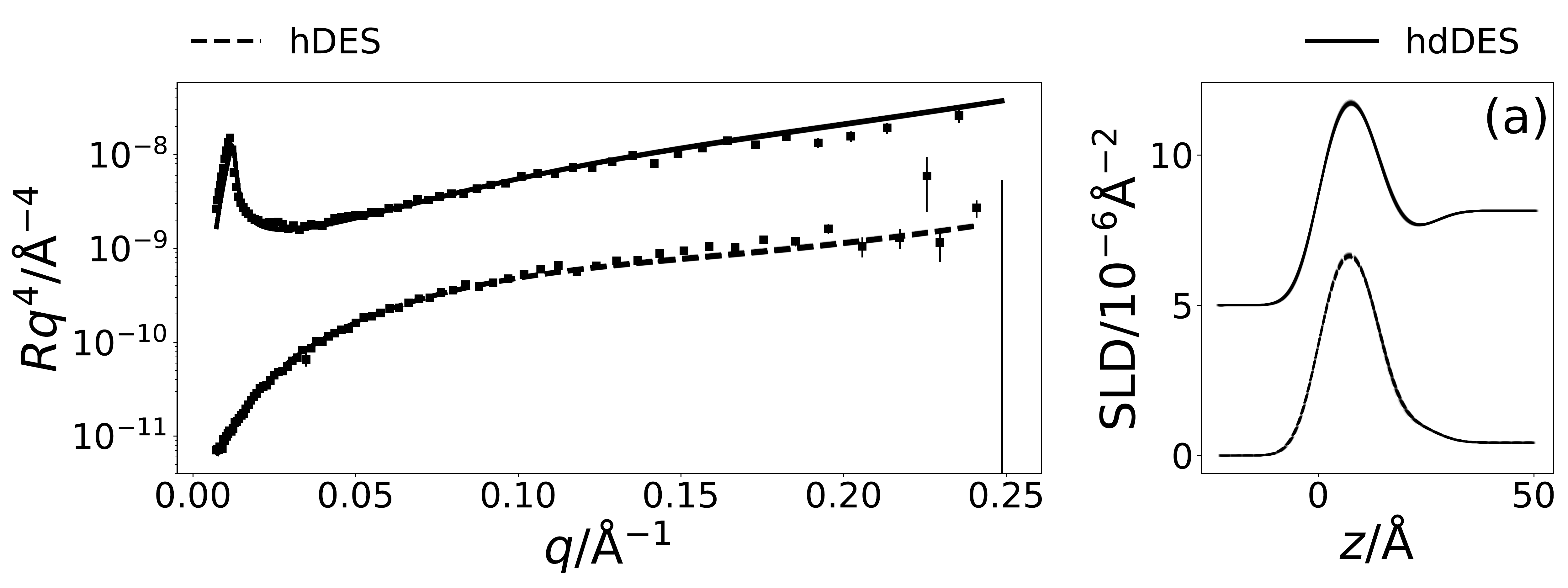}
    \includegraphics[width=0.45\textwidth]{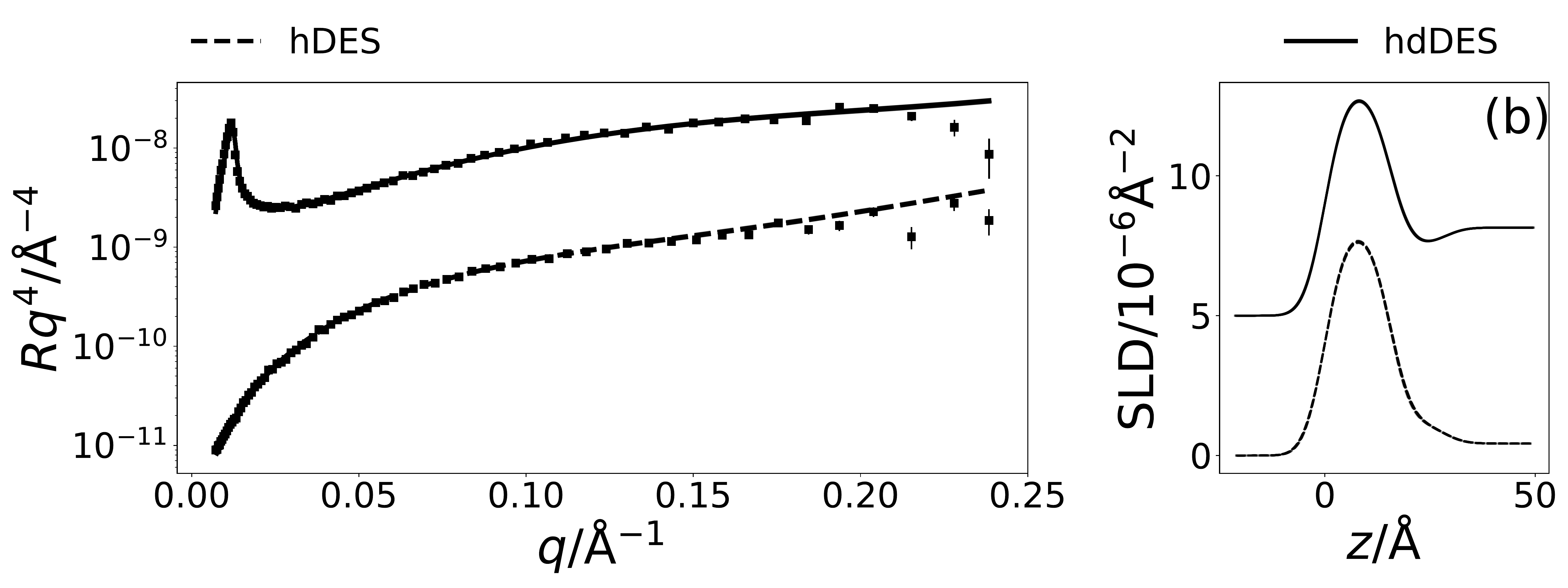}
    \caption{\small The NR and SLD profiles at a surface pressure of
  20 mNm$^{-1}$ for two contrasts, the solid line identifing the hdDES contrast and the dashed line the hDES contrast; (a) DMPC, (b) DPPC. The NR profiles have been offset in the $y$-axis by an order of magnitude and SLD profiles offset in the $y$-axis by \SI{5e-6}{\per\square\angstrom}, for clarity.}
    \label{fig:neutron}
\end{figure}

Following the initial structural determination of the monolayer by XRR, NR was used to confirm the structure and show the applicability of the chemically-consistent model for DPPC and DMPC.
The resulting NR profiles and associated SLD profiles, at a surface pressure of 20 mNm$^{-1}$ are given in Figure \ref{fig:neutron} (the other NR and SLD profiles can be found in Figure \ref{fig:sineutron} of the ESI).
Table \ref{tab:neutron} gives details of the varied parameters at each surface pressure as well as $\phi_h$ as determined from NR.
\begin{table*}
  \caption{\label{tab:neutron} The best-fit values, and associated 95 \%
  confidence intervals for the varying parameters in the co-refined NR models.
  The values of $\phi_h$ were found using Eqn. \ref{equ:phih}.}
  \begin{ruledtabular}
    \begin{tabular*}{\textwidth}{ccccc}
    Lipid & \multicolumn{2}{c}{d$_{54}$-DMPC} & \multicolumn{2}{c}{d$_{62}$-DPPC} \\
    SP/mNm$^{-1}$ & 20 & 25 & 15 & 20 \\
    \hline
    $\sigma_{t,h,s}$/\si{\angstrom} & \input{dmpc_rough20_neutron.txt} & \input{dmpc_rough25_neutron.txt} & \input{dppc_rough15_neutron.txt} & \input{dppc_rough20_neutron.txt} \\
    $d_t$/\si{\angstrom} & \input{dmpc_tail20_neutron.txt} & \input{dmpc_tail25_neutron.txt} & \input{dppc_tail15_neutron.txt} & \input{dppc_tail20_neutron.txt} \\
    \hline
    $\phi_h$/$\times10^{-2}$ & \input{dmpc_solh20_neutron.txt} & \input{dmpc_solh25_neutron.txt} & \input{dppc_solh15_neutron.txt} & \input{dppc_solh20_neutron.txt} \\
    \end{tabular*}
  \end{ruledtabular}
\end{table*}
\subsection{Effect of compression on monolayer thickness}
From Tables \ref{tab:liptab} and \ref{tab:neutron}, we can see that, as expected and as found in previous work \cite{mohwald_phospholipid_1990,vaknin_structural_1991}, the thickness of the tail layer increases as the number of carbon atoms in the tail chain increases.
Furthermore, the thickness of the tail layers in these monolayers appears to agree well with values found for water-analogues; \input{dmpc_tail30.txt}\AA\ at 30 mN/m in DES compared with $d_t=15.8$~\si{\angstrom} at \SI{30}{\milli\newton\per\meter} in water for DMPC \cite{johnson_structure_1991}, and \input{dppc_tail30.txt}~\si{\angstrom} at \SI{30}{\milli\newton\per\meter} in DES compared with $d_t=16.7$~\si{\angstrom} at \SI{40}{\milli\newton\per\meter} in water for DPPC \cite{helm_phospholipid_1987}.

\begin{figure}
    \centering
    \includegraphics[width=0.45\textwidth]{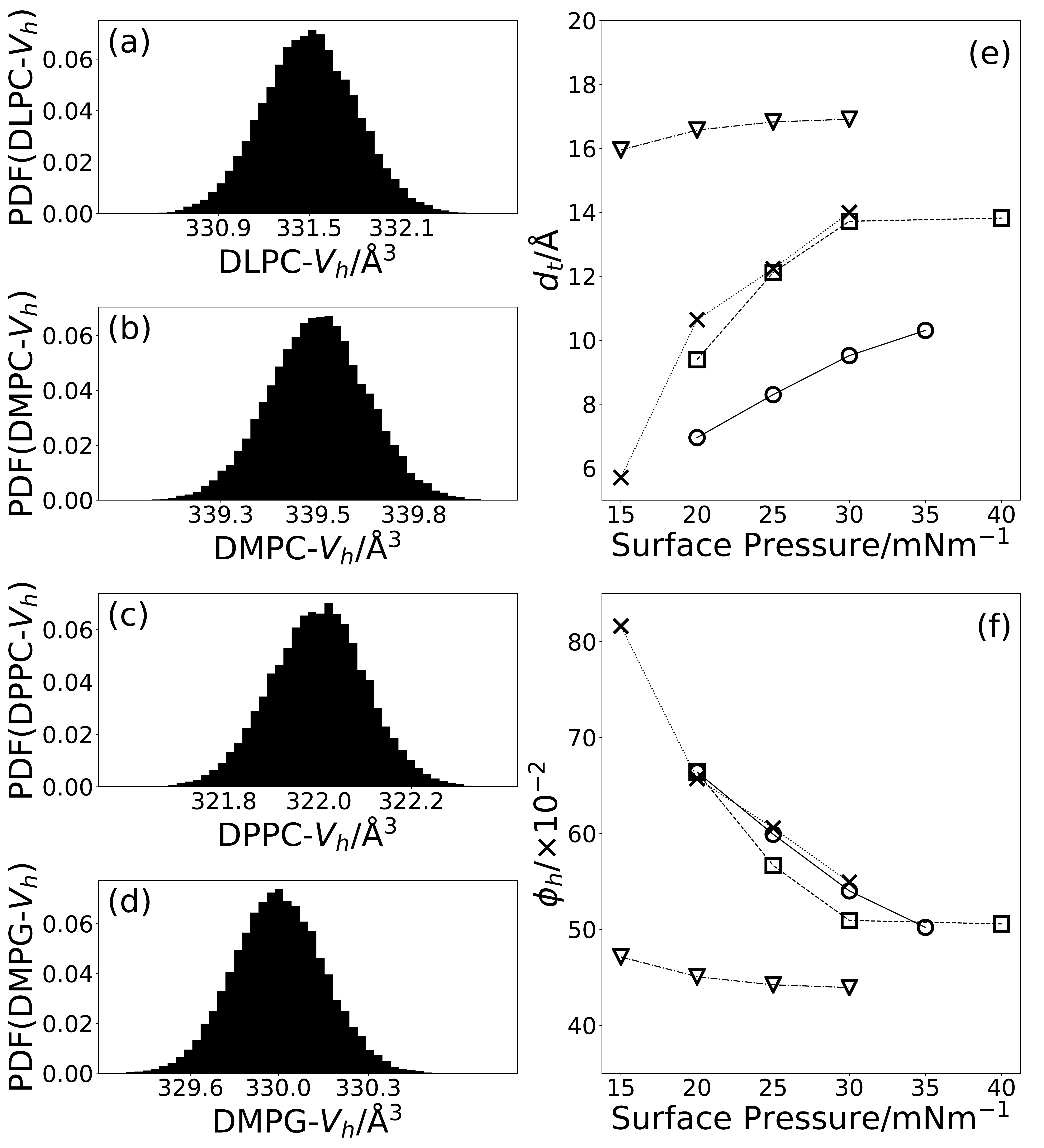}
    \caption{\small The PDFs of the head volume for each of the four lipids; (a) DLPC, (b) DMPC, (c) DPPC, and (d) DMPG, and the variation of the tail layer thickness (e) and volume fraction of solvent in the head layer (f); DLPC (circles), DMPC (squares), DPPC (triangles), and DMPG (crosses), trend lines have been added between the points to guide the eye.}
    \label{fig:lipresults}
\end{figure}

The variation of the tail layer thickness in the models with surface pressure is given for each lipid in Figure \ref{fig:lipresults}(e).
As is commonly observed for lipid monolayers at the air-water interface, an increase in surface pressure results in an increase in the thickness of the tail layer.
The variation of this parameter differs for each of the lipids.
For DLPC and DMPC a gradual increase in the thickness is observed at the surface pressures measured here.
For DMPC and DPPC however, after a gradual increase in thickness between \SIlist{15;25}{\milli\newton\per\meter}, a plateau is reached.
This is indicative of the formation of a compacted monolayer where further compression does not produce further changes in the thickness of the tail layer.
This behaviour has been seen for DMPC and DPPC monolayers, where the formation of the condensed phases is reached at high surface concentrations \cite{mohwald_phospholipid_1990}.

\subsection{Effect of compression on solvent concentration}
In Figure \ref{fig:lipresults}(f), it is clear that for all four lipids, as the surface pressure is increased there is a corresponding decrease in the percentage solvent present in the lipid head layer.
This can be rationalised by considering that when the surface pressure is increased, the free volume available to the solvent between the lipid head components reduces forcing the solvent out of the lipid head layer and into the bulk.
A similar effect has been observed when increasing the surface pressure from \SI{11}{\milli\newton\per\meter} to \SI{31}{\milli\newton\per\meter} for a DMPC/DMPG monolayer at the air-water interface \cite{bayerl_specular_1990}.

\subsection{Effect of compression on the lipid tail component volumes}
It can be seen by comparing Tables \ref{tab:water} and \ref{tab:liptab} that the volumes of the lipid tails are significantly lower in the current measurements than found previously, by other techniques.
It is unlikely that this is a result of the DES subphase, due to the hydrophobic nature of the lipid tails.
However, this reduction has been shown previously \cite{campbell_structure_2018}, where it was rationalised by the compaction of the monolayer at elevated surface pressure.
In that work, the optimal value of the tail component volume for DPPC was found to be \SI{772}{\angstrom\cubed} at a surface pressure of $35$ mNm$^{-1}$, this agrees well with the value of \input{dppc_vt.txt}~\si{\angstrom\cubed} found in this work at surface pressures of \SIlist[list-units = single]{15;20;25;30}{\milli\newton\per\meter}.

In this work, a single tail component volume was fitted to each lipid for all four surface pressures that were measured.
This is based on the assumption, that at all four surface pressures, the lipids adopt the same phase and therefore any variation in the structure with surface pressure would manifest only as a change in the tail thickness.
It is clear when comparing Tables \ref{tab:water} and \ref{tab:liptab} that some of the tail component volumes are also reduced in the current XRR measurements compared to those determined previously.
The reduction was found to be between \SIrange{8}{12}{\percent} for DPPC, DMPC and DLPC when compared with literature sources at \SIrange{24}{30}{\celsius}, this is in good agreement with the maximum compression percentage of \SI{15}{\percent} noted by Small and coworkers \cite{small_lateral_1984}.
DMPG shows a small increase in the tail volume relative to the literature value quoted at a lower temperature.
Notably, this value is similar to that found in this work for DMPC, which has the same tail structure and suggests that our results are at least self-consistent.

\subsection{Solvent effect on lipid head component volumes}
Figure \ref{fig:lipresults} shows the PDFs determined for the head component volume for each of the four lipids.
The three lipids with the PC head component are consistent, giving values of $\sim330$ \AA$^3$, regardless of tail component.
This agrees well with the values found for the same head component in water, shown in Table \ref{tab:water}.
Interestingly, the component volume for the PG head is similar to that for the PC head with a value of \input{dmpg_vh.txt}~\si{\angstrom\cubed}.
The PG head component volume in water, from either DMPG using differential vibrating tube densimetry \cite{pan_molecular_2012} or POPG using molecular dynamics simulations \cite{kucerka_scattering_2012}, is noticeably smaller.
This indicates that there may be some effect arising from the solvation in choline chloride:glycerol causing an apparent increase in the PG component volume when compared with water.
However, this has only been shown for a single PG-lipid at the air–DES interface.

The major difference between the two head groups of the lipids is that the PG is present as a sodium salt, whereas the PC is zwitterionic.
When in solution the anionic PG head is expected to associate with cations in solution, as it does in water \cite{grigoriev_effect_1999} where such interactions depended on a variety of factors including the ionic strength.
In the case of a DES, the environment is inherently ionic and therefore the interaction of an anionic lipid head may be more complex.
As well as interacting with the sodium, the head is likely to interact with the choline cations, similar to behaviour reported previously for surfactant micelles \cite{sanchez-fernandez_counterion_2018}.
The extent of interaction with each of the cations is unclear, but regardless it seems likely that the solvation of the PG head is improved in the DES relative to water.
This better solvation would explain the apparent increase in the volume of the PG head since it would result in a swelling of this group through its strong interactions with the solvent.
In the case of PC, the proximity of a local cation within the molecule results in the same folding of the head group seen in water because this interaction is less transient than the equivalent interactions with the solvent.

\subsection{Refinement of neutron reflectometry}
The ability to fit the NR data, as shown in Figure \ref{fig:neutron} indicates that the value found for the head component volume is consistent between the pair of measurements for the same system.
It is clear, that again stable monolayers of the lipids are forming at the air-DES interface, and that the component volumes determined from XRR measurements are robust enough to be used in the modelling of NR data.
Furthermore, the trends observed with increasing surface pressure in the XRR models, pertaining to the increasing tail thickness and decreasing solvent concentration in the head components are consistent with that found in the NR models.

\subsection{Interparameter correlations}
\begin{figure}
    \centering
    \includegraphics[width=0.48\textwidth]{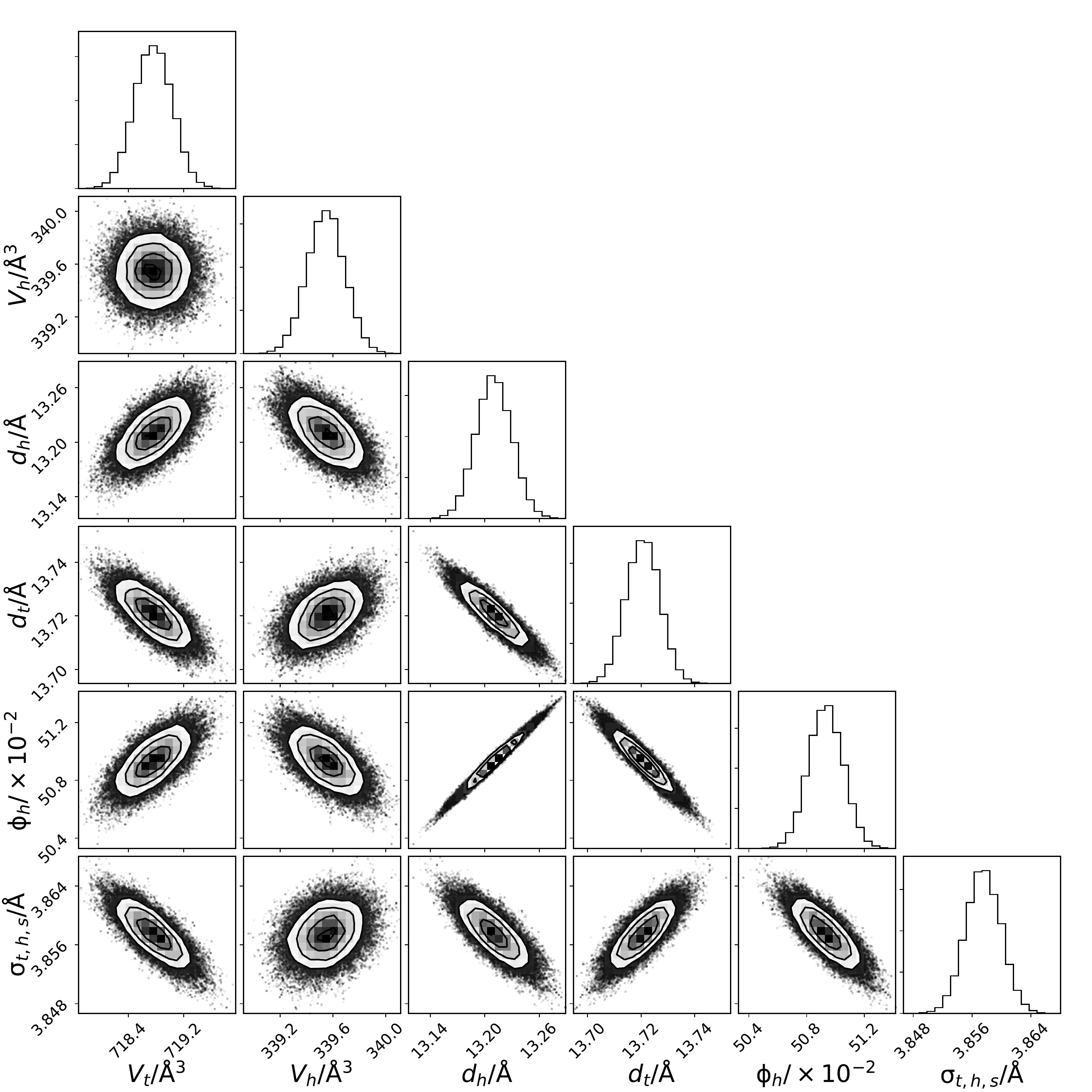}
    \caption{The multi-parameter PDFs for the chemically-consistent model of DMPC XRR data at \SI{30}{\milli\newton\per\meter}.}
    \label{fig:dmpcpdfs}
\end{figure}

The use of Bayesian inference and MCMC sampling allowed for the probing of the probability distribution function for each parameter individually.
However, it also enables the pairwise inter-parameter PDFs to be investigated, an example for DMPC at \SI{30}{\milli\newton\per\meter} is shown in Figure \ref{fig:dmpcpdfs} (similar plots for the other XRR datasets and the NR datasets are available in Figures \ref{fig:dlpc1} - \ref{fig:dppcn2} of the ESI).
These two-dimensional PDFs give important information about the correlations that are present between the parameters of the chemically-consistent model.
The less circular in nature that the 2D PDF is, the greater the correlation that is present between the parameters, e.g. a north-east/south-westerly skewed PDF indicates a positive correlation (where an increase in one parameter correlates an increase in the other) while a north-west/south-easterly skewed PDF indicates a negative correlation (an increase in one correlates with a decrease in the other).

Substantial correlations are present in the parameters fitted to the XRR datasets, indicating important uncertainty that must be considered.
In particular, as can be seen in Figure \ref{fig:dmpcpdfs}, there is a positive correlation between the lipid head thickness, $d_h$, and the solvent concentration in the head layer, $\phi_h$.
This correlation can be rationalised as a result of the SLD of the solvent and the head layer (which is \SI{\sim50}{\percent} solvated) being similar, and therefore the boundary between the two is not easy to define.
Such correlations are unavoidable without considering many neutron contrasts of the lipid and solvent, due to the solvophilic nature of the lipid heads.
Another important correlation is that between the head thickness and the tail thickness, $d_t$, again this is due to the lack of a well-defined boundary between the head and tail layers.
This is partially driven by the interfacial roughness that is present between the layers, with the correlation being more pronounced for phospholipids with shorter tails (e.g. there is a greater correlation for DLPC than DPPC).

\section{Conclusions}
For the first time, stable phosphocholine and phosphatidylglycerol lipid monolayers have been observed and characterised on an ionic solvent surface.
Until the emergence of ionic liquids and DES, only a limited number of molecular solvents exhibited the ability to promote self-assembly and, to the best of our knowledge, only water and formamide among those had demonstrated the formation of phospholipid monolayers at the air-liquid interface.

A physically and chemically constrained modelling approach and Bayesian analysis method was used to rationalise these measurements showing that the structures are remarkably similar at the air-DES interface to those previously observed at the air-water interface.
This has the important implication that DES, therefore, offer the possibility of performing studies of model membranes in the absence of water.
Such applications may include fundamental investigations of phospholipid monolayers in extreme environments (total or partial absence of water, cryogenic temperatures), protein-membrane interactions and development of new technologies for drug delivery.
However, the PG component did show a significant difference; having a larger head component volume than observed for the same system in water.
This suggests that the transfer of lipids to a DES is not just a simple substitution of the subphase.
In this specific case, we have proposed an explanation based on the dissociation of the PG head component salt and the subsequent interaction with the DES.
Finally, the use of MCMC sampling of the reflectometry model parameter space gives insight into the correlations present in XRR data modelling, that should be considered in work of this nature.

The ability to determine the head component volume was facilitated by access to easy to use, open-source software that allowed for the straightforward use of a custom, chemically-consistent model within the analysis of the XRR and NR measurements.
Furthermore, to our knowledge, this work presents the first use of a chemically-consistent parameterisation to co-refine XRR measurements at different surface concentrations.

\section{Author Contributions}
T.A., A.J.J., A.S.-F. and K.J.E. were involved in the experimental design and conducted the experiments along with R.A.C..

A.R.M. developed the chemically-consistent model and analysed the data, with input from S.C.P, A.S.-F., T.A., and R.A.C..

A.R.M., A.S.-F. and T.A. wrote the manuscript, with input from all authors.

\acknowledgments{
The authors thank Andrew Nelson for useful discussions \& assistance with the refnx software.
Additionally, we thank Oliver Hammond for insightful discussions regarding the effect of the DES.
A.R.M. is grateful to the University of Bath and Diamond Light Source for
co-funding a studentship (Studentship Number STU0149).
Thanks also to the European Spallation Source and the University of Bath
Alumni Fund for supporting A.S.-F.
We also thank Diamond Light Source (Experiment number SI10546-1) and
Institut Laue-Langevin (DOI:
\href{http://doi.org/10.5291/ILL-DATA.9-13-612}{10.5291/ILL-DATA.9-13-612})
for the awarded beamtime.}

\bibliography{bibi}

\end{document}

%% file: dlpc_rough30.txt
$4.17\pm{0.02}$

%% file: dmpc_rough30.txt
$3.86\pm{0.00}$

%% file: dppc_rough30.txt
$4.90\pm{0.00}$

%% file: dmpg_rough30.txt
$4.44\pm{0.01}$

%% file: dlpc_tail30.txt
$9.52^{+0.03}_{-0.04}$

%% file: dmpc_tail30.txt
$13.72\pm{0.01}$

%% file: dppc_tail30.txt
$16.91\pm{0.01}$

%% file: dmpg_tail30.txt
$13.99^{+0.01}_{-0.01}$

%% file: dlpc_vt.txt
$624.92\pm{3.51}$

%% file: dmpc_vt.txt
$718.76\pm{0.52}$

%% file: dppc_vt.txt
$765.29^{+0.37}_{-0.38}$

%% file: dmpg_vt.txt
$734.01\pm{0.62}$

%% file: dlpc_vh.txt
$331.48\pm{0.58}$

%% file: dmpc_vh.txt
$339.55\pm{0.28}$

%% file: dppc_vh.txt
$322.01\pm{0.24}$

%% file: dmpg_vh.txt
$329.95^{+0.32}_{-0.33}$

%% file: dlpc_head.txt
$10.98^{+0.13}_{-0.12}$

%% file: dmpc_head.txt
$13.21\pm{0.04}$

%% file: dppc_head.txt
$12.69\pm{0.03}$

%% file: dmpg_head.txt
$13.95\pm{0.04}$

%% file: dlpc_solh30.txt
$54.03^{+1.04}_{-0.95}$

%% file: dmpc_solh30.txt
$50.93\pm{0.23}$

%% file: dppc_solh30.txt
$43.94\pm{0.22}$

%% file: dmpg_solh30.txt
$54.92\pm{0.20}$

%% file: dmpc_rough20_neutron.txt
$4.42\pm{0.16}$

%% file: dmpc_rough25_neutron.txt
$3.31^{+0.01}_{-0.02}$

%% file: dppc_rough15_neutron.txt
$4.27\pm{0.17}$

%% file: dppc_rough20_neutron.txt
$3.98\pm{0.10}$

%% file: dmpc_tail20_neutron.txt
$13.98\pm{0.15}$

%% file: dmpc_tail25_neutron.txt
$17.97^{+0.05}_{-0.01}$

%% file: dppc_tail15_neutron.txt
$12.32\pm{0.13}$

%% file: dppc_tail20_neutron.txt
$15.56\pm{0.10}$

%% file: dmpc_solh20_neutron.txt
$50.00^{+0.54}_{-0.54}$

%% file: dmpc_solh25_neutron.txt
$35.72^{+0.04}_{-0.16}$

%% file: dppc_solh15_neutron.txt
$59.16^{+0.43}_{-0.43}$

%% file: dppc_solh20_neutron.txt
$48.40^{+0.33}_{-0.33}$